\documentclass[11pt,twoside]{paper}

\usepackage{asp2010}
\usepackage{graphicx}
\usepackage{bm}

\resetcounters

\begin{document}

\newcommand{\be}{\begin{equation}}
\newcommand{\ee}{\end{equation}}
\newcommand{\bdm}{\begin{displaymath}}
\newcommand{\edm}{\end{displaymath}}
\newcommand{\bea}{\begin{eqnarray}}
\newcommand{\eea}{\end{eqnarray}}

\title{High-order schemes for non-ideal $3+1$ GRMHD: 
a study of the kinematic dynamo process in accretion tori}  

\author{L. Del Zanna$^{123}$, M. Bugli$^{14}$, N. Bucciantini$^{23}$}   
\affil{$^1$Dipartimento di Fisica e Astronomia, Universit\`a di Firenze, Italy}     
\affil{$^2$INAF - Osservatorio Astrofisico di Arcetri, Firenze, Italy}
\affil{$^3$INFN - Sezione di Firenze, Italy}    
\affil{$^4$Max-Planck-Institut f\"ur Astrophysik, Garching, Germany}    

\begin{abstract}
We present the first astrophysical application of the ECHO code in its recent version 
supplemented by a generalized Ohm law, namely a kinematic study of dynamo effects 
in thick accretion disks. High-order \emph{implicit-explicit} Runge-Kutta
time-stepping routines are implemented and validated within $3+1$ \emph{General Relativistic
MagnetoHydroDynamics} (GRMHD).
The scheme is applied to a differentially rotating torus orbiting a Kerr black hole, 
where the mean-field dynamo process leads to strong amplification of seed magnetic fields. 
We show that the interplay between the toroidal and poloidal components occurs qualitatively 
in the same fashion as in the Sun, butterfly diagrams are reproduced, and a typical time-scale 
for the field evolution is found, depending on the dynamo and resistivity numbers,
which could explain periodicities as observed in several accreting systems.
\end{abstract}


\section{Introduction}

Large-scale, ordered magnetic fields are believed to play a crucial role to activate
the mechanisms responsible for the observed high-energy emission, from active
galactic nuclei to gamma-ray bursts. However, it is still not clear which is the
precise origin of such fields. 
While the process of collapse to the compact objects which are believed to power these sources
is certainly able to amplify any existing frozen-in field, the question of its origin just shifts
to the progenitors. Turbulent motions and small-scale instabilities, such as the 
\emph{Magneto-Rotational Instability} (MRI) in accretion disks or the Tayler instability
in proto-neutron stars, are indeed good candidates to enhance also the level
of turbulent magnetic fields, but the resulting configurations will be highly tangled
and thus not appropriate to generate the required large-scale fields.

A process naturally able to amplify ordered magnetic fields is that of \emph{dynamo}.
Here we shall consider the so-called \emph{mean-field} dynamo processes \citep{Moffatt:1978},
i.e. when small-scale turbulent (correlated) motions, invariably arising in
astrophysical plasmas with very high fluid and magnetic Reynolds numbers, give rise
to an effective ponderomotive force, along $\bf{B}$, in the Ohm law. This is capable in turn
to increase any initial seed field when coupled to the Faraday evolution equation.
The resistive-dynamo Ohm law can be written, for classical MHD, as (here $c\to 1$,
$4\pi\to 1$)
\be
\bm{E}^\prime \equiv \bm{E} + \bm{\varv}\times\bm{B} = \eta \bm{J} + \xi \bm{B},
\label{eq:ohm_classic}
\ee
where $\eta$ is the coefficient of resistivity (due to collisions and to the mean-field effects),
and $\xi$ is the mean-field dynamo coefficient (in the literature $\alpha = - \xi$ is most 
commonly employed) . The latter term is precisely responsible for 
the dynamo process, leading to the generation of exponentially growing modes in the 
kinematical phase, that is when the feedback of the increasing field on the other quantities 
can still be neglected. Differential rotation also plays a role in the dynamo process, and when
both effects are at work we refer to an $\alpha\Omega$ dynamo.

While in classical MHD the above relation simply defines the electric field, which
is a derived quantity of $\bm{\varv}$ and $\bm{B}$ alone (since $\bm{J}=\nabla\times\bm{B}$), 
in the relativistic case the displacement current in the Maxwell equations cannot be neglected, 
and the evolution equation for the electric field must be solved as well.
This difficulty was avoided in previous relativistic studies of kinetic dynamo in accretion disks
\citep{Khanna:1996a,Brandenburg:1996}. To our knowledge the first generalization to full General
Relativity and its formulations for $3+1$ numerical relativity can be found in \cite{Bucciantini:2013}
(BDZ from now on), where also a systematic validation of the second-order numerical scheme,
within the ECHO \citep{Del-Zanna:2007} and X-ECHO \citep{Bucciantini:2011a} codes, is presented.
Here we briefly summarize the formalism and we improve the numerical method employed
to higher-order. Moreover, we present, as a first numerical application, a study of the kinematic 
dynamo action in thick accretion tori around Kerr black holes. More details and additional simulations
will be presented in a forthcoming paper \citep{Bugli:2014}.

\section{Basic equations}

The fully covariant formulation for a resistive plasma with dynamo action, first proposed
in BDZ, is a relation for the quantities measured in the local frame comoving with the fluid
4-velocity $u^\mu$, namely
\be
e^{\mu}=\eta j^{\mu}+\xi b^{\mu},
\label{eq:ohm_covariant}
\ee
to be compared with  Eq.~(\ref{eq:ohm_classic}), where $e^\mu$, $b^\mu$, and $j^\mu$ 
are the electric field, magnetic field, and current as measured by the observer moving
with $u^\mu$.
When written in the $3+1$ formalism \citep{Gourgoulhon:2012},
the Maxwell equations become
\be
\gamma^{-1/2}\partial_{t}\left(\gamma^{1/2}\bm{B}\right)+
\bm{\nabla}\times(\alpha\bm{E}+\bm{\beta}\times\bm{B})  =   0, 
\quad (\bm{\nabla}\cdot\bm{B}=0),
\ee
\be
\gamma^{-1/2}\partial_{t}\left(\gamma^{1/2}\bm{E}\right)+
\bm{\nabla}\times(-\alpha\bm{B}+\bm{\beta}\times\bm{E})  =  - (\alpha\bm{J}-q\bm{\beta}) , 
\quad (\bm{\nabla}\cdot\bm{E}=q),
\ee
where the electromagnetic fields $\bm{E}$ and $\bm{B}$ are those measured by the
\emph{Eulerian observer}, $\alpha$ is the \emph{lapse function}, $\bm{\beta}$ the
spatial \emph{shift vector}, $\gamma_{ij}$ the 3-metric with determinant $\gamma$,
$q$ is the charge density and $\bm{J}$ the usual conduction current.
In the $3+1$ language, Ohm's law (\ref{eq:ohm_covariant}) translates into
\be
\Gamma[\bm{E}+\bm{v}\times\bm{B}-(\bm{E}\cdot\bm{v})\bm{v}]=
\eta(\bm{J}-q\bm{v})+\xi\Gamma[\bm{B}-\bm{v}\times\bm{E}-(\bm{B}\cdot\bm{v})\bm{v}],
\label{eq:ohm}
\ee
here $\bm{v}$ is the Eulerian velocity and $\Gamma = (1-v^2)^{-1/2}$ the
usual Lorentz factor,
from which one can finally compute $\bm{J}$, so that the new equation for $\bm{E}$ is
\begin{eqnarray}
&\gamma^{-1/2}\partial_{t} (\gamma^{1/2}\bm{E} )
+ \bm{\nabla}\times(-\alpha\bm{B}+\bm{\beta}\times\bm{E}) + 
(\alpha\bm{v}-\bm{\beta})\bm{\nabla}\cdot\bm{E}  = \nonumber \\  &
 - \alpha\Gamma\,\eta^{-1} \{ [\bm{E}+\bm{v}\times\bm{B}-(\bm{E}\cdot\bm{v})\bm{v}] 
- \xi [\bm{B}-\bm{v}\times\bm{E}-(\bm{B}\cdot\bm{v})\bm{v}]\},
\label{eq:stiff}
\end{eqnarray}
reducing to $\bm{E} = - \bm{v} \times \bm{B}$ for $\eta=\xi = 0$ and to Eq.~(\ref{eq:ohm_classic})
in the non-relativistic case.

\section{Numerical settings and convergence tests}

In typical astrophysical situations the resistivity coefficient is very small and thus
Eq.~(\ref{eq:stiff}) is a \emph{stiff} equation: terms $\propto\eta^{-1}$ can evolve on timescales 
$\tau_{\eta}\ll\tau_{h}$, where the latter is the hyperbolic (fluid) timescale. We have then to solve,
even for the simple kinematic case, a system of hyperbolic partial differential equations
combined with stiff relaxation equations, that can be rewritten in the form
\begin{eqnarray}
\bm{X}=\gamma^{1/2}\bm{E}: \quad \partial_{t}\bm{X} & = & \bm{Q}_{X}(\bm{X},\bm{Y})+\bm{R}_X(\bm{X},\bm{Y}), \\
\bm{Y}=\gamma^{1/2}\bm{B}: \quad \partial_{t}\bm{Y} & = & \bm{Q}_{Y}(\bm{X},\bm{Y}),
\end{eqnarray}
where $\bm{Q}_{X,Y}$ are the non-stiff terms, whereas
$\bm{R}_X$ are the stiff terms $\propto \eta^{-1}$
requiring some form of implicit treatment in order to preserve numerical stability.
To solve this system, we here extend the numerical methods in BDZ
(up to second order in time and space) to higher orders. For spatial integration we use
the full set of high-order shock-capturing methods implemented in the ECHO code,
whereas for time integration we use the \emph{IMplicit-EXplicit}
(IMEX) Runge-Kutta methods developed by \cite{Pareschi:2005} and first introduced
to resistive (special) relativistic MHD by \cite{Palenzuela:2009}.

\begin{figure}
\begin{center}
\includegraphics[scale=0.34]{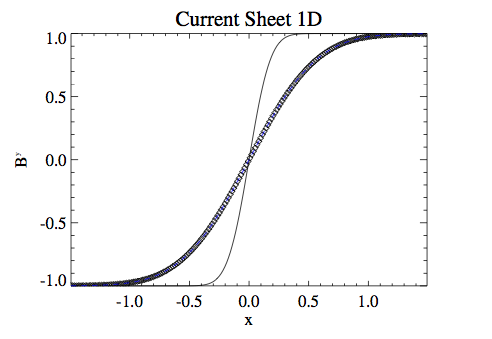}
\includegraphics[scale=0.38]{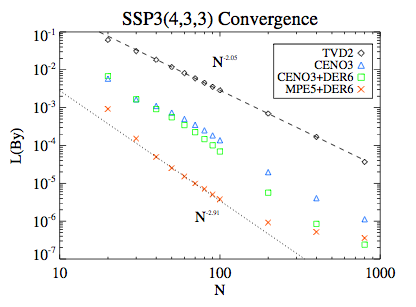}
\includegraphics[scale=0.36]{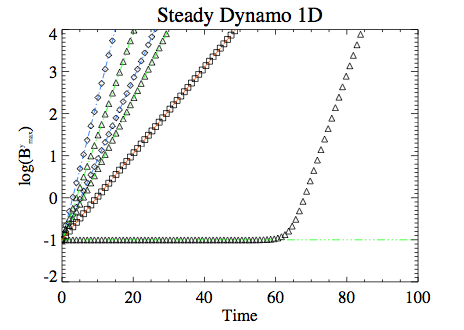}
\includegraphics[scale=0.33]{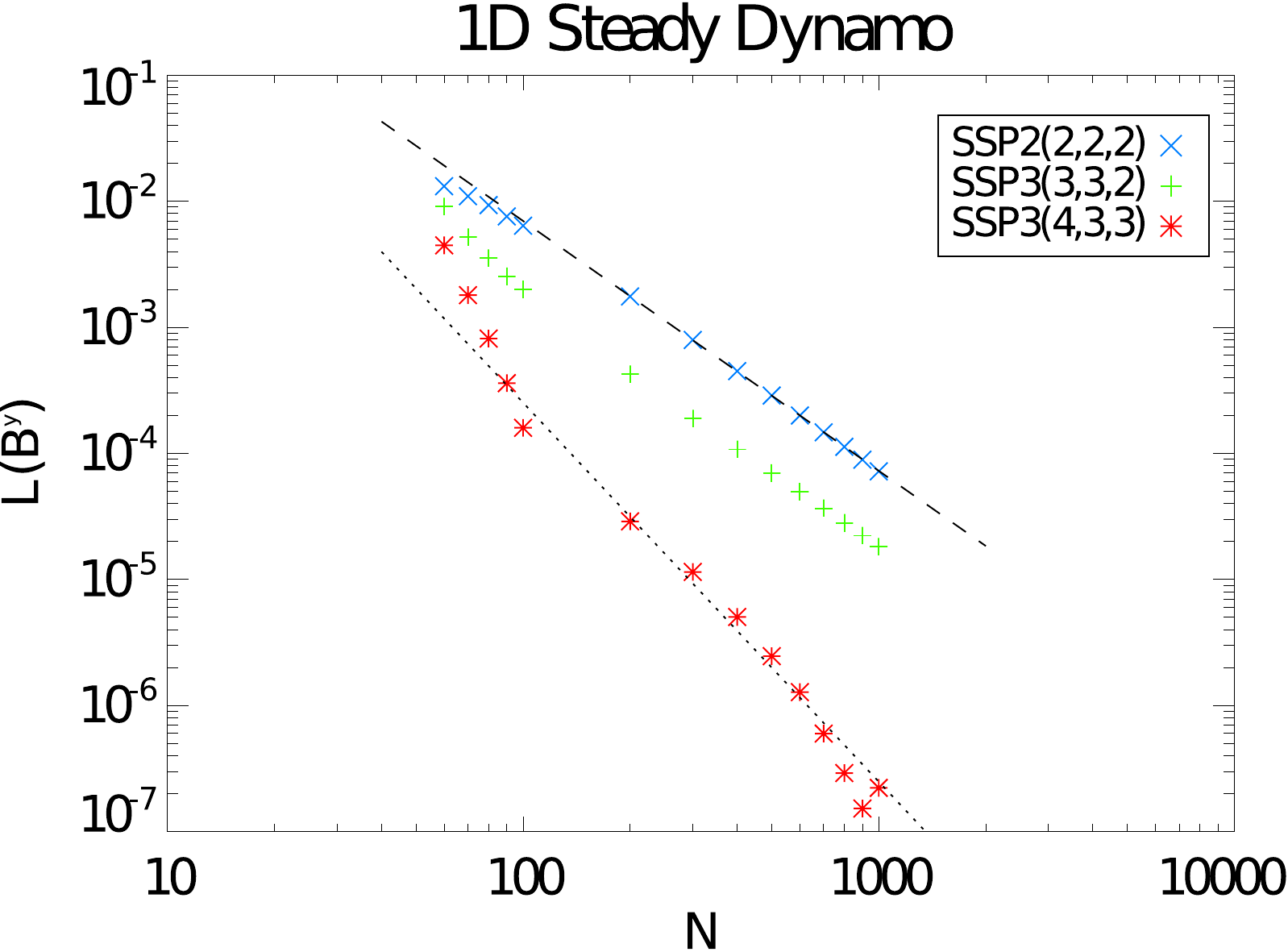}
\caption{Top panels: convergence study for the 1D current sheet test.
Bottom panels: convergence study for the 1D steady dynamo test.}
\label{fig:test}
\end{center}
\end{figure}

In Fig.~\ref{fig:test} we show a couple of 1D numerical tests in Minkowski spacetime,
already described in BDZ. The first one is the resistive diffusion of a smooth current sheet,
followed from $t=1$ to $t=10$. Here $\eta=0.01$, $\xi=0$, and $p\gg B^2$ to reduce compressible
effects (this test is made by coupling the Maxwell equations to the full set of GRMHD, we
do not consider the kinematic case). In this limit the system remains non-relativistic 
and the analytical solution is
\be
B^y(x,t) = B_0\mathrm{erf}\left(\frac{x}{2\sqrt{\eta t}}\right),
\ee
like for the classical diffusion equation. In the top row, left panel, we show the analytical
solution at the initial and final times, and the numerical one at $t=10$, in the case $B_0=0.01$. 
In the right panel we show the $L_1$ errors on $B^y$ (comparing with a run at extremely high 
resolution) as a function of the grid points number $N$. When we use the third-order SSP3(4,3,3)
IMEX method, an overall second or third order space-time accuracy is correctly
achieved, depending on the spatial reconstruction method employed. The best
performing scheme is REC-MPE5 combined to DER-E6 \citep{Del-Zanna:2007}.

The second test concerns exponentially growing dynamo modes in a static, magnetized
background. The analytical solution, for a given wave vector $k$, is
\be
B^y(x,t) = B_0\mathrm{exp} (\gamma t) \mathrm{cos} (kx), \quad
\gamma = (2\eta)^{-1}[\sqrt{1+4\eta k (\xi - \eta k)}-1],
\ee
where $\gamma$ is the dynamo growth factor. In Fig.~\ref{fig:test} we show in the bottom
row, left panel, the theoretical and measured amplification of $B^y(t)$ for various
combinations of $\eta$ (diamonds for $\eta=0.05$, triangles for $\eta=0.1$, squares for $\eta=0.25$), 
$\xi=0.5$, and $k$. The correspondence is always matched. The case of $\eta=0.1$ and $k=5$
should not have growing modes, due to the stabilizing effect of resistivity, but at late times 
($t\simeq 60$) small-scale fluctuations due to truncation errors triggers the fastest possible
growing modes anyway. In the right panel we show the $L_1$ errors on $B^y$ for various
choices of the IMEX schemes (and for REC-MPE5, DER-E6). The nominal second or third order
is achieved.

\section{Kinematic dynamo in accretion tori}

We choose as the background equilibrium the differentially rotating thick disk
in Kerr metric and Boyer-Lindquist coordinates described in \cite{Del-Zanna:2007}.
We select the maximally rotating case with $a=0.99$, resulting in an orbital period of $P_c=76.5$ 
(in code units) at the center. 
The domain is $r\in [2.5,25]$ (logarithmically stretched) and 
$\theta\in [\pi/4-0.2,3\pi/4+0.2]$, with a numerical resolution as low as $120\times 120$
due to the long-term runs needed to capture the full dynamics. The IMEX-RK is
SSP3(4,3,3), thus third-order in time, and we use REC-MPE5 and DER-E6 for spatial
reconstruction and derivation (thus fifth order in space for smooth profiles).

In the kinematic case the dynamics is basically determined by the two
\emph{dynamo numbers}, corresponding to the respective importance of the $\alpha$
and $\Omega$ effects with respect to diffusion. We take
$C_{\xi}=\xi R / \eta\geq 1$ and $C_\Omega=\Delta\Omega R^2/\eta \gg 1$,
where $R$ is the radial extension of the thick disk (and vertical too),
$\Delta\Omega$ the difference in the angular velocity between the disk center and
the inner radius. The choice of these two numbers determine the values of $\xi$
and $\eta$ at the disk center. The coefficients are then modulated within the disk
as the square root of the mass density and proportional to it, respectively, while
we take $\xi=0$ and $\eta=10^{-5}$ in the low-density atmosphere
(co-rotating to minimize shear flow numerical dissipation). 
The dynamo coefficient is further multiplied by $\cos\theta$ because we want 
to obtain a combination of the $\alpha\Omega$
dynamo effect which is symmetrical with respect to the equator.

\begin{figure}
\begin{center}
\includegraphics[height=8cm]{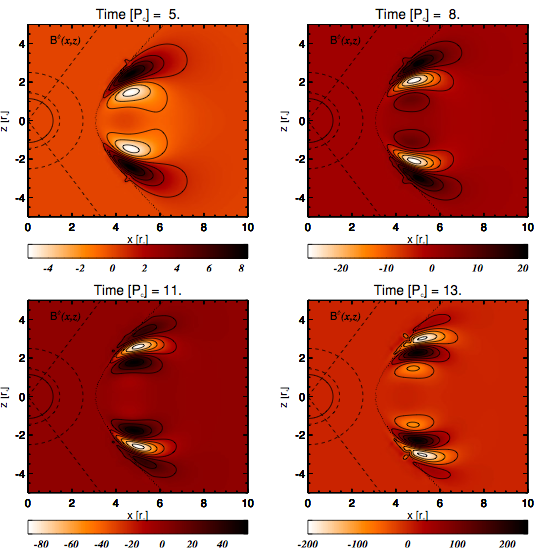}
\caption{Dynamo evolution of the toroidal field for $\xi=\eta=10^{-3}$.}
\label{fig:Bphi} 
\end{center}
\end{figure}

\begin{figure}[b]
\begin{center}
\includegraphics[scale=0.25]{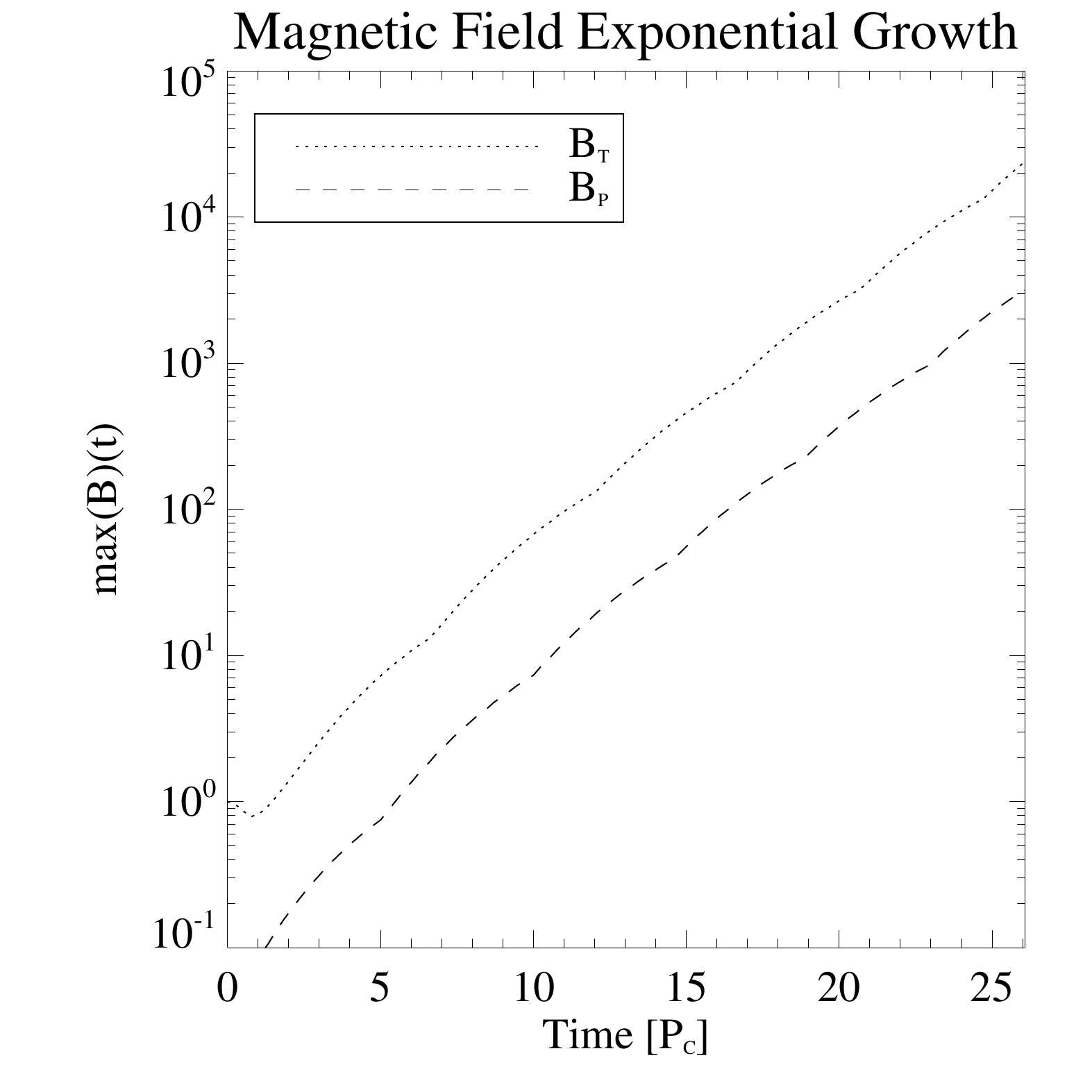}
\includegraphics[scale=0.25]{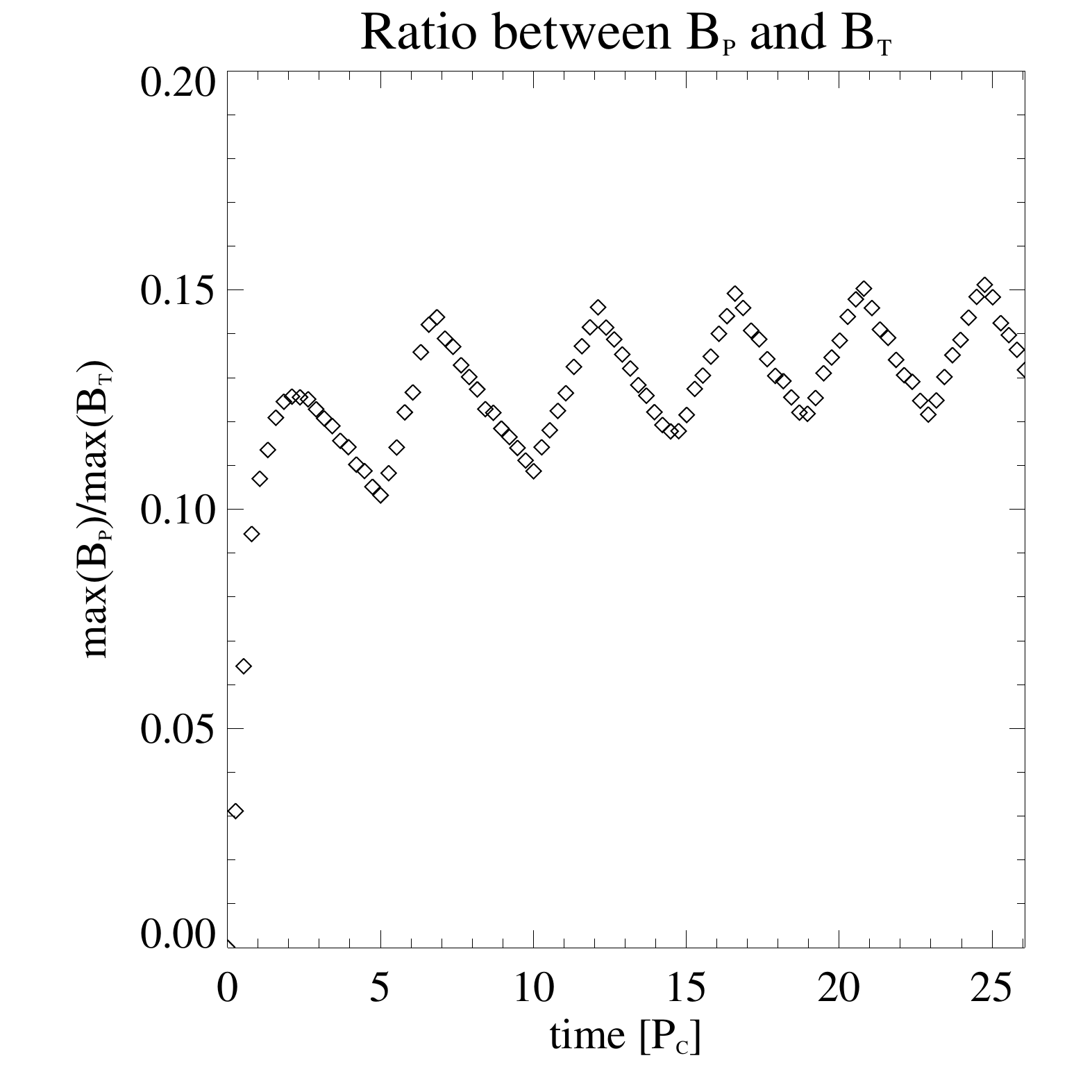}
\caption{Growth of the toroidal and poloidal components and $B_P/B_T$ ratio.}
\label{fig:Bratio}
\end{center}
\end{figure}

The initial seed field ($B\sim 10^{-5}$) can be taken either as purely toroidal 
or purely poloidal, in the latter case it is derived from a potential $A_\phi$
proportional to the square of the local pressure.
In Fig.~\ref{fig:Bphi} we show a simulation with $\xi=\eta=10^{-3}$ (maximum values
at the disk center), corresponding to $C_\xi = 5$ and $C_\Omega = 400$, with
an initial toroidal field. This is shown as a function of time (in units of $P_c$) and
we can clearly see the periodical formation and migration of structures from the
center to higher latitudes, always confined within the disk.
In Fig.~\ref{fig:Bratio} we show the maximum value of the toroidal and poloidal
orthonormal components of the field in the disk and their ratio, as a function of time.
We observe the exponential increase of both components and the saturation
of the ratio around a value $B_P/B_T \simeq 0.13$, while the oscillating behavior is
simply due to a difference in phase between the dynamo modes for the two
components.

\begin{figure}
\begin{center}
\includegraphics[height=4.5cm]{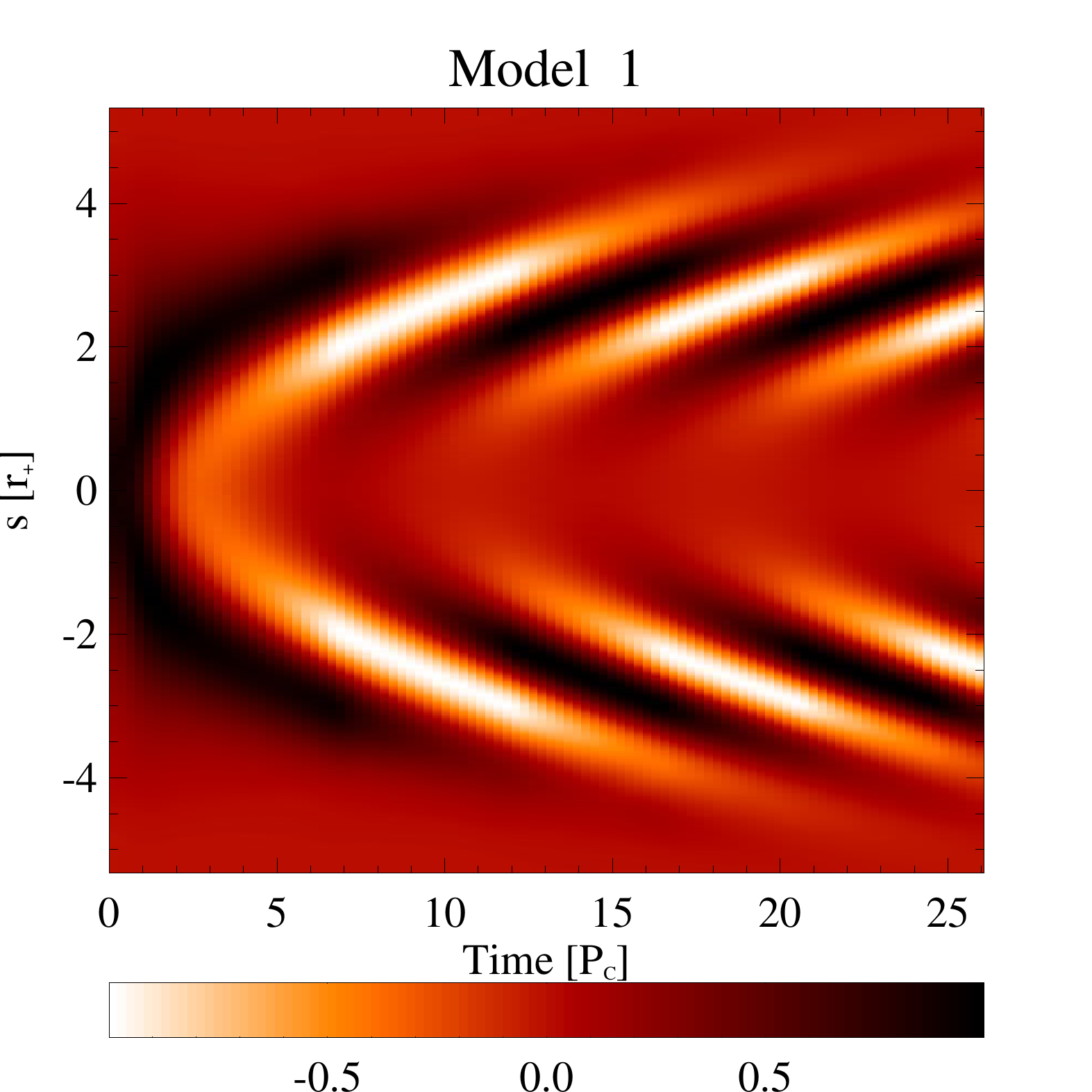}
\includegraphics[height=4.5cm]{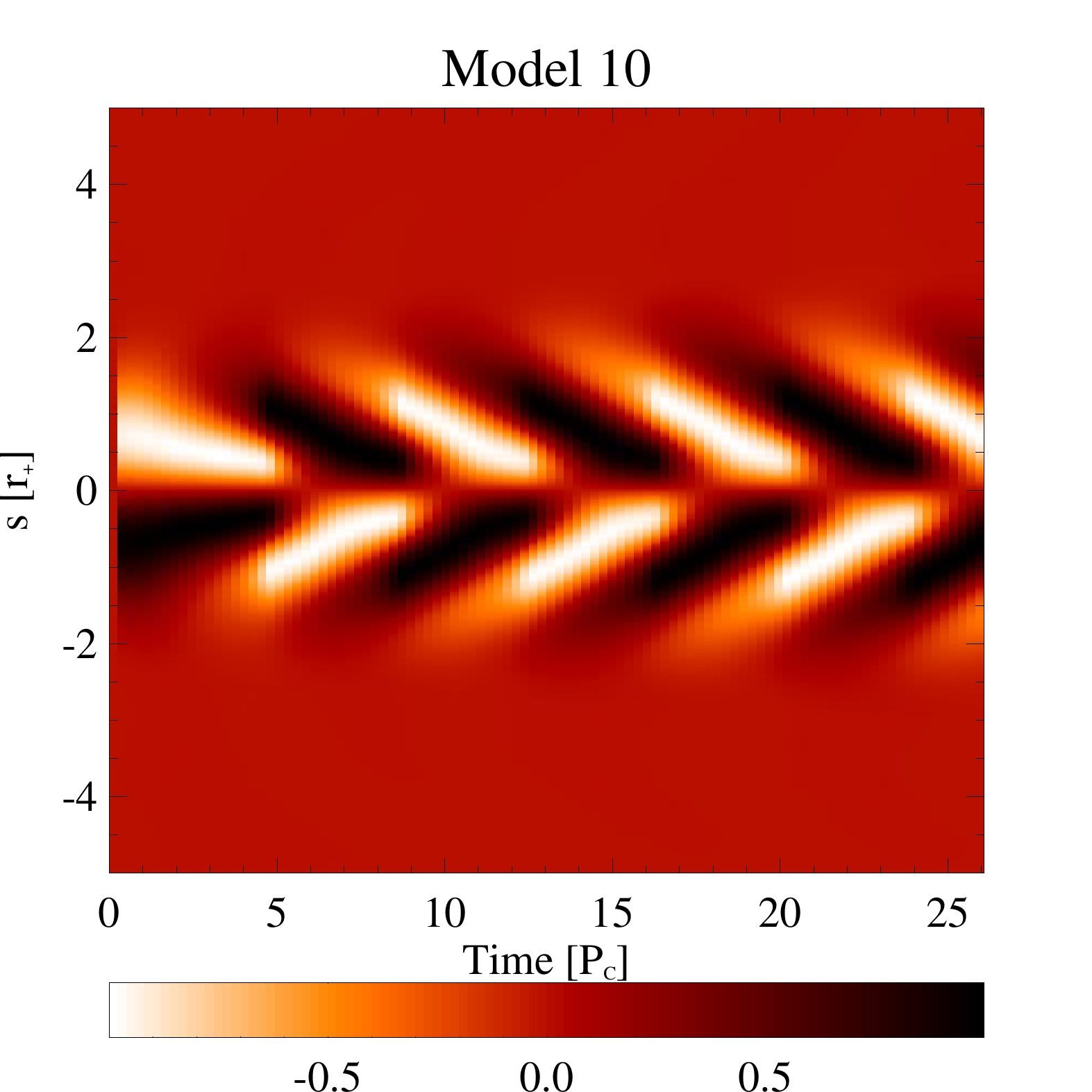}
\caption{Butterfly diagrams for $\xi=\pm 10^{-3}$ and $\eta=10^{-3}$.}
\end{center}
\label{fig:butterfly}
\end{figure}

The periodical generation and migration of structures resembles very much
what happens in the Sun, where the $\alpha\Omega$ is believed to be
responsible for the 11-years sun-spots cycle. This is usually tracked in the so-called
\emph{butterfly diagrams}, with periodical field reversal and migration \emph{towards} 
the solar equator.
In Fig.~\ref{fig:butterfly} we report the position of the peaks along an appropriate
new coordinate $s$ as a function of time. In the left panel the run previously
described is shown, with migration from the equator to higher latitudes ($\xi>0$
in the northern hemisphere). In the right panel the sign of $\xi$ has been reversed
(and the field initialized to a purely poloidal one), to better reproduce the solar
situation (we recall that $\alpha = - \xi$ in solar dynamo applications). 

To conclude, in this paper we have shown the first implementation of high-order
IMEX-RK methods for the resistive-dynamo non-ideal GRMHD version of the ECHO code. 
Preliminary runs of the kinematic $\alpha\Omega$ dynamo effect
in thick accretion tori around maximally rotating Kerr black holes have ben performed
as an astrophysical test case. Similarly to the solar cycle, we have found periodical
generation and migration of magnetic structures, with exponential growth of
the maximum value of the field, expected to be quenched by non-linear feedback when
the full set of GRMHD will be solved. The (half) cycle of the dynamo process
is seen to be around $6-7$ orbital periods, but this value can change according
to the parameters chosen. In any case the turbulence properties can set an
additional timescale, through the mean-field dynamo effect, unrelated to
the larger-scale quantities. This property may help to explain periodical modulations
in the accreting process and in thus the source luminosity itself \citep[e.g.][]{Gilfanov:2010}.

\bibliographystyle{asp2010}

\end{document}